\documentclass[sigconf]{acmart}

\usepackage{booktabs}
\usepackage{amsmath}
\usepackage{multirow}
\usepackage{tikz}
\usetikzlibrary{positioning, arrows.meta, shapes.geometric, calc}
\usepackage{fontspec}
\newfontfamily\arabicfont{Amiri-Regular.ttf}[Path=./, Script=Arabic]
\newcommand{\rtl}[1]{\beginR #1\endR}

\setcopyright{none}                
\renewcommand\footnotetextcopyrightpermission[1]{} 

\acmConference[RecSys '26]{Proceedings of the 20th ACM Conference
on Recommender Systems}{September 28--October 2, 2026}
{Minneapolis, MN, USA}
\copyrightyear{2026}
\acmYear{2026}

\begin{document}

\title{Multilingual Semantic Retrieval for Apple Music Search}
\titlenote{Accepted to the Industry Track of the 20th ACM Conference on Recommender Systems (RecSys 2026). This is the authors' accepted manuscript; the final version will appear in the ACM Digital Library.}

\author{Vishalaksh Aggarwal}
\authornote{Equal contribution.}
\affiliation{\institution{Apple}\city{Cupertino}\state{CA}\country{USA}}
\email{vishalaksh@apple.com}

\author{Kevin Sebastian}
\authornotemark[2]
\affiliation{\institution{Apple}\city{Cupertino}\state{CA}\country{USA}}
\email{k_chittinappilyse@apple.com}

\author{Vivek Kanojiya}
\affiliation{\institution{Apple}\city{Cupertino}\state{CA}\country{USA}}
\email{vkanojiya@apple.com}

\author{Leo Le}
\affiliation{\institution{Apple}\city{Cupertino}\state{CA}\country{USA}}
\email{leo_le@apple.com}

\author{Nick Tucey}
\affiliation{\institution{Apple}\city{Cupertino}\state{CA}\country{USA}}
\email{ntucey@apple.com}

\author{Santosh Shankar}
\affiliation{\institution{Apple}\city{Cupertino}\state{CA}\country{USA}}
\email{santosh\_shankar@apple.com}

\begin{abstract}
Apple Music serves listeners across 150+ storefronts in dozens of languages,
with a catalog that grows by hundreds of thousands of new tracks daily. At this scale, search
recall on misspelled, transliterated, and cross-lingual queries becomes a
dominant driver of session quality, particularly for tail queries that account
for the majority of unique queries. We present a multilingual semantic
retrieval system built on a 305M-parameter Siamese bi-encoder fine-tuned from
GTE-multilingual-base~\cite{li2023towards} with curriculum-scheduled
multi-objective training. The model is integrated into the search stack via a
hybrid retrieval architecture that blends dense nearest-neighbor results with
the existing token-based index using quantile distribution matching, enabling
deployment without retraining downstream rankers. Offline, the model achieves
a 69\% relative improvement in Hit@10 over GTE-multilingual-base. In a worldwide online A/B test, the system delivers a 2.28\%
relative conversion-rate (CR) lift overall, an 86\% reduction in the no-result
rate, and gains across every storefront with no observed regressions. The
improvement is concentrated where it is needed most: tail queries see a 7.93\%
relative CR lift, compared with 0.89\% for mid-frequency queries and 0.14\%
for head queries---evidence that semantic retrieval improves recall on hard
queries without disturbing well-served popular ones. To our knowledge, this is
one of the largest search-quality improvements deployed on the platform.
\end{abstract}



\maketitle

\section{Introduction}
\label{sec:introduction}
 
Apple Music users search across a catalog that spans content types such as songs, albums, artists, playlists, and stations. The existing retrieval system uses a token-based index with language-specific analyzers: query tokens are matched against normalized document metadata, and a downstream ranker re-scores the lexical candidates. This works well for head queries, where lexical overlap is high and engagement signals are abundant. However, our analysis of search sessions indicates that, among sessions with poor result quality, the majority are recall failures rather than ranking errors, with misspellings and other surface-form mismatches leading to poor or no recall being the dominant failure modes.
 
These recall failures are concentrated in \emph{tail queries}---queries with low historical frequency that account for 83\% of unique queries and roughly one-third of all search sessions. The conversion rate\footnote{The fraction of search sessions resulting in a meaningful engagement such as plays exceeding a minimum duration threshold, add-to-library actions, or shares.} is lower for tail queries than for head queries, where token-matching pipelines struggle because lexical overlap between query and intended item is minimal or absent. For example, a user searching for ``gnarly by cat's eye'' (a misspelling of Gnarly by KATSEYE) finds nothing; a Persian-speaking user searching ``{\arabicfont\rtl{مارتیک بهار}}'' for the song ``Bahar'' by Martik sees zero matches when translation coverage for that item is incomplete in their language; a search for content not yet in our catalog returns empty rather than a closely related item; and a functional query like ``party music from the 90s''---or its Spanish equivalent ``Música de fiesta de los años 90''---shares little vocabulary with the editorial playlists that best satisfy it (``'90s Hits Essentials'' and ``'90s Latin Essentials'', respectively). Each of these failure modes is amplified by the multilingual setting, where the same catalog item may be sought under many surface forms across dozens of languages, scripts, and transliterations.
 
We address these challenges through three contributions:
 
\begin{enumerate}
    \item \textbf{A multilingual bi-encoder for music search.} We introduce
    ELISE, a 305M-parameter Siamese model fine-tuned from
    GTE-multilingual-base~\cite{li2023towards} with
    curriculum-scheduled, multi-objective training over Query--Item and
    Query--Query tasks, combined with data strategies including
    storefront-level frequency capping, synthetic misspelling augmentation, and
    world-knowledge injection---each targeted at a measured failure mode of
    music search.
 
    \item \textbf{Hybrid integration without ranker retraining.} A
    quantile distribution matching scheme that fuses dense ANN scores with
    the existing token-based index while preserving the original text-match
    score distribution, enabling deployment without retraining downstream
    rankers.
 
    \item \textbf{Tail-targeted, regression-free deployment at worldwide
    scale.} A worldwide A/B test demonstrating a 2.28\% relative CR lift
    overall, with the gain sharply concentrated in the tail (+7.93\% relative
    CR) and head queries effectively unchanged (+0.14\%)---evidence that
    semantic retrieval can augment a mature lexical pipeline without head regressions. The system
    also delivers an 86\% reduction in the no-result rate and positive gains
    across every storefront with no regressions.
\end{enumerate}
\section{Related Work}
\label{sec:related}

\paragraph{Dense Retrieval for Entity Search.}
Dense passage retrieval using dual-encoder architectures trained with contrastive objectives~\cite{karpukhin2020dense} has been widely adopted, with improvements from hard negative mining~\cite{gao2021condenser, xiong2021approximate} and heterogeneous evaluation~\cite{thakur2021beir}. A growing line of work applies dense retrieval to \emph{structured entity search} at industry scale: Facebook social search over people, pages, and groups~\cite{huang2020embeddingfb}, and product retrieval at Amazon~\cite{nigam2019semantic}, Taobao~\cite{li2021taobao}, and Walmart~\cite{magnani2022semantic}, where challenges of vocabulary mismatch, noisy queries, and tail-query recall parallel those in music search.

\paragraph{Multilingual Text Embeddings.}
Multilingual BERT~\cite{devlin2019bert}, XLM-R~\cite{conneau2020unsupervised}, and more recently E5~\cite{wang2022text} and GTE~\cite{li2023towards} have advanced cross-lingual text representations. M3-Embedding~\cite{chen2024m3embedding} further extends this line with unified dense, sparse, and multi-vector retrieval across 100+ languages. The MTEB benchmark~\cite{muennighoff2023mteb} provides standardized evaluation for comparing these models.

\paragraph{Hybrid Retrieval.}
Combining sparse lexical retrieval with dense semantic retrieval has shown consistent improvements across domains~\cite{luan2021sparse,kuzi2020leveraging,chen2022out}. A key challenge is score calibration between the two sources.

\paragraph{Music Information Retrieval.}
Music search presents unique challenges including artist name ambiguity, transliteration across scripts, and the prevalence of misspellings in song and artist names~\cite{schedl2014music}. Recent work has explored joint audio-text embeddings: MuLan~\cite{huang2022mulan} learns to align music audio with natural-language descriptions, CLAP~\cite{elizalde2023clap} extends contrastive language-audio pretraining more broadly, and follow-on work targets contrastive music-language learning~\cite{manco2022contrastive} and universal text-to-music retrieval~\cite{doh2023toward}. While these approaches target content-based audio retrieval, our work addresses a complementary problem: text-based semantic search over structured catalog metadata at production scale.

\paragraph{Curriculum Learning.}
Curriculum learning~\cite{bengio2009curriculum} structures the training process to progress from simpler to more complex objectives, typically by ordering training examples by difficulty. Our work applies a related principle at the loss-function level rather than to data ordering (Section~\ref{sec:curriculum}).
\section{System Overview}
\label{sec:system}

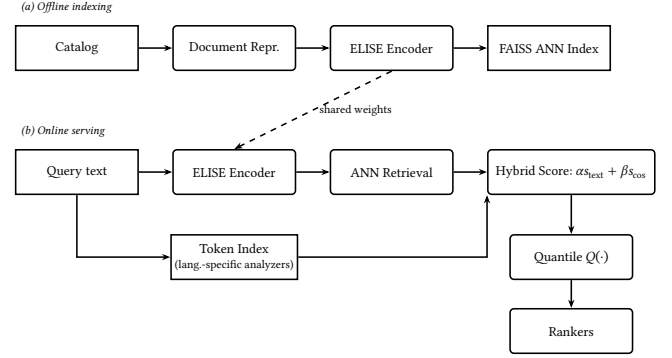
\begin{figure}[t]
\centering
\resizebox{\columnwidth}{!}{%
\begin{tikzpicture}[
  font=\footnotesize,
  >={Stealth[length=1.5mm]},
  proc/.style   = {rectangle, rounded corners=2pt, draw, thick,
                   minimum height=8mm, minimum width=22mm,
                   align=center, inner sep=2pt},
  store/.style  = {rectangle, draw, thick,
                   minimum height=8mm, minimum width=22mm,
                   align=center, inner sep=2pt},
  lane/.style   = {font=\scriptsize\itshape},
  arr/.style    = {->, thick},
  shared/.style = {->, thick, dashed},
  node distance = 5mm and 6mm,
]

\node[store]                       (catalog) {Catalog};
\node[proc, right=of catalog]      (docrep)  {Document Repr.};
\node[proc, right=of docrep]       (enc1)    {ELISE Encoder};
\node[store, right=of enc1]        (ann)     {FAISS ANN Index};

\draw[arr] (catalog) -- (docrep);
\draw[arr] (docrep)  -- (enc1);
\draw[arr] (enc1)    -- (ann);

\node[lane, anchor=south west]
  at ([yshift=1mm]catalog.north west) {(a) Offline indexing};

\node[store, below=14mm of catalog]                 (query)  {Query text};
\node[proc, right=of query]                         (enc2)   {ELISE Encoder};
\node[proc, right=of enc2]                          (anns)   {ANN Retrieval};
\node[proc, right=of anns, minimum width=30mm]      (hyb)    {Hybrid Score: $\alpha s_{\text{text}} + \beta s_{\cos}$};
\node[store, below=7mm of enc2]                     (tok)    {Token Index\\\scriptsize(lang.-specific analyzers)};
\node[proc, below=7mm of hyb]                       (qmap)   {Quantile $Q(\cdot)$};
\node[proc, below=of qmap]                          (rank)   {Rankers};

\draw[arr] (query) -- (enc2);
\draw[arr] (enc2)  -- (anns);
\draw[arr] (anns)  -- (hyb);
\draw[arr] (query.south) |- (tok.west);
\draw[arr] (tok.east) -| (hyb.south west);
\draw[arr] (hyb)   -- (qmap);
\draw[arr] (qmap)  -- (rank);

\node[lane, anchor=south west]
  at ([yshift=1mm]query.north west) {(b) Online serving};

\draw[shared] (enc1.south) -- node[right, font=\scriptsize] {shared weights} (enc2.north);

\end{tikzpicture}%
}
\caption{System architecture. Offline (a): catalog entities are encoded by the ELISE text encoder and indexed in a FAISS ANN store. Online (b): a query is encoded by the same shared-weight encoder (dashed arrow) for ANN retrieval and tokenized for the existing sparse index; the two candidate sets are fused via hybrid scoring and quantile calibration $Q(\cdot)$ before the rankers.}
\label{fig:architecture}
\end{figure}

Figure~\ref{fig:architecture} illustrates the end-to-end system, which combines two retrieval branches: a new \emph{dense (semantic) path} and the existing \emph{sparse (lexical) path}. The dense path uses an ELISE encoder and a single ANN index spanning all supported locales; the sparse path uses a token-based index built with language-specific analyzers. Each branch comprises an offline indexing pipeline and an online serving stage, detailed below.

\paragraph{Offline Pipeline.}
Catalog entities (e.g., songs, albums, artists, playlists, radio stations) are encoded into 256-dimensional embeddings by the ELISE text encoder applied to structured document representations (Section~\ref{sec:doc-rep}). These embeddings are indexed into a FAISS~\cite{johnson2019billion} ANN index with composite quantization (Section~\ref{sec:ann}).

\paragraph{Online Serving.}
At query time, the dense path encodes the raw query with the ELISE encoder and retrieves candidates from the ANN index, while the sparse path tokenizes the query with language-specific analyzers and retrieves candidates from the existing token-based index. The union of results is then fused and calibrated before being passed to the downstream rankers (Section~\ref{sec:blending}). Latency of the dense path is reported in Section~\ref{sec:ab}.

\section{ELISE Model}
\label{sec:model}

\subsection{Architecture}
\label{sec:architecture}

ELISE is a Siamese bi-encoder~\cite{reimers2019sentence} initialized from GTE-multilingual-base~\cite{li2023towards}, a 305M-parameter transformer pretrained on 75+ languages. The architecture uses shared weights for both the query and document towers. We extract the \texttt{[CLS]} token representation from the final hidden layer and retain the first 256 dimensions, leveraging the backbone's Matryoshka Representation Learning (MRL)~\cite{kusupati2022matryoshka} training, which ensures that any prefix of the full embedding remains a valid representation. We then apply L2 normalization, and similarity is computed as the inner product between normalized embeddings (equivalent to cosine similarity).

We disable RoPE NTK scaling~\cite{liu2024scaling} in the GTE backbone since our input sequences (max 512 tokens for documents, 32 for queries) are well within the base model's position range, and NTK scaling introduces unnecessary approximation at shorter lengths.

\subsection{Document Representation}
\label{sec:doc-rep}

Rather than encoding raw metadata fields independently, we construct a structured text template that linearizes all available entity metadata into a single input string. For a song entity, the template takes the form:

\smallskip
{\ttfamily\sloppy\noindent Document:: Entity Type: song; Title: Knockin' On Heaven's Door; Artist: Bob Dylan; Genre: Roots Rock, Classic Rock, Folk-Rock, Singer/Songwriter; Release Information: published long ago, published in the 70s, year: 1973; Popularity: level~4 (somewhat high); Album Name: Bob Dylan's Greatest Hits, Vol.~3; Top Queries: bib dillan, bov dulan, robert allen zimmerman; Language: English; Lyrics: <lyrics snippet>\par}
\smallskip

Queries are wrapped with a distinct \texttt{Query::} prefix, mirroring \texttt{Document::}. Since the two towers share weights, this prefix lets self-attention condition on the input role and adapt accordingly---short, noisy lexical patterns for queries versus structured metadata for documents---without separate per-tower parameters.

Key design choices for individual fields include:

\begin{itemize}
    \item \textbf{Popularity bucketing:} Continuous popularity scores are discretized into six natural-language levels (``low'' through ``very high'') to provide a stable textual signal rather than a noisy numeric one.
    \item \textbf{Release information:} Epoch timestamps are converted to human-readable strings capturing recency (e.g., ``recently published,'' ``published long ago''), decade, and year.
    \item \textbf{Lyrics:} Full lyrics exceed the context window. To optimize for search retrieval, we index a representative subset of the text by selecting the first $k$ lines, the $m$ most frequently repeated lines, and the last $x$ lines.
    \item \textbf{Top queries:} To bridge the vocabulary gap between how users search and how items are described, we inject past engaged queries as additional document features. These are non-metadata queries that previously led to engagement with the item, denoised via edit-distance and Jaccard similarity filtering to remove near-duplicates of existing metadata. The retained queries often capture signals that other fields cannot---extreme misspellings beyond our edit-distance threshold (e.g., ``bib dillan'') and world-knowledge associations (e.g., ``robert allen zimmerman'' for Bob Dylan)---giving the model exposure to how users actually formulate searches.
\end{itemize}

The same template structure is adapted for other entity types such as albums, artists, playlists, and radio stations by including entity-specific fields (e.g., curator name for playlists, popular song names for artists).

\subsection{Training Objectives}
\label{sec:objectives}

ELISE is trained with two complementary objectives:

\paragraph{Query-Item (QI) Objective.}
The primary task learns to map search queries to their relevant catalog items (used interchangeably with ``document'' and ``entity'' throughout). Training pairs are sampled from worldwide query logs where the query led to a successful engagement (play, add-to-library, or other conversion action), and therefore span the full distribution of real user queries: natural-language, navigational, genre/mood, similarity, and misspelling queries alike. The QI loss receives a weight of $\lambda_{\text{QI}}$ in the final objective.

\paragraph{Query-Query (QQ) Objective.}
The secondary task aligns failed queries with successful reformulations from the same session, directly targeting the misspelling failure mode. The QQ loss receives a weight of $\lambda_{\text{QQ}}$.

The final training loss is:
\begin{equation}
    \mathcal{L} = \lambda_{\text{QI}} \cdot \mathcal{L}_{\text{QI}} + \lambda_{\text{QQ}} \cdot \mathcal{L}_{\text{QQ}}.
\end{equation}

\subsection{Curriculum Training}
\label{sec:curriculum}

Both objectives follow a two-stage schedule that transitions from broad contrastive alignment to focused margin-based discrimination:

\paragraph{Stage 1: InfoNCE with In-Batch Negatives.}
For the QI task, we use a composite loss combining symmetric InfoNCE~\cite{oord2018representation} over in-batch negatives with standard InfoNCE over uniformly sampled catalog negatives. In-batch negatives reflect query-log frequency, so popular items appear as negatives at the same rate as they do as positives---preventing the model from over-recommending popular items, which would otherwise dominate the training signal. Uniformly sampled catalog negatives complement this with broader coverage of the long tail~\cite{zhang2020dpsr}. The composite loss is:
\begin{equation}
    \mathcal{L}_{\text{QI}}^{(1)} = \frac{1}{2}\left(\mathcal{L}_{\text{sym}} + \mathcal{L}_{\text{neg}}\right),
\end{equation}
where $\mathcal{L}_{\text{sym}} = \frac{1}{2}(\mathcal{L}_{q\to d} + \mathcal{L}_{d\to q})$ is the bidirectional InfoNCE over in-batch negatives: each direction is a cross-entropy over the in-batch query-document similarity matrix, with $\mathcal{L}_{q\to d}$ treating each query as the anchor against all in-batch documents and $\mathcal{L}_{d\to q}$ reversing query and document roles. Both use temperature $\tau = 0.07$. $\mathcal{L}_{\text{neg}}$ is the standard InfoNCE loss against uniformly sampled catalog negatives.

For the QQ task, we use symmetric InfoNCE only, dropping the catalog-negatives term: each failed query is paired with its successful rewrite as the positive, with all other rewrites in the batch serving as negatives.

\paragraph{Stage 2: Pairwise Hinge with Hardest-in-Batch Negatives.}
At epoch 0.5, we switch both objectives to a pairwise hinge loss that focuses on the single hardest negative per query within the batch:
\begin{equation}
    \mathcal{L}_{\text{hinge}} = \frac{1}{B}\sum_{i=1}^{B} \max\!\left(0,\; m - s(q_i, d_i^+) + \max_{j \neq i} s(q_i, d_j^-)\right),
\end{equation}
where $m = 0.4$ is the margin and $s(\cdot,\cdot)$ denotes cosine similarity. This two-stage schedule applies curriculum learning~\cite{bengio2009curriculum} at the loss-function level: in early training, the gradient receives a softmax-weighted contribution from every in-batch negative, establishing global embedding structure, while later training backprops through only the single hardest negative, sharpening decision boundaries for confusable pairs. The switch point (epoch 0.5) was selected by evaluating validation MRR at regular intervals during training and transitioning when this curve plateaued. The loss weights $\lambda_{\text{QI}}$, $\lambda_{\text{QQ}}$, the temperature $\tau$, and the hinge margin $m$ were selected via grid search over a small set of candidates, optimizing Hit@10 on the validation set (final values: $\lambda_{\text{QI}}=0.8$, $\lambda_{\text{QQ}}=0.2$). Section~\ref{sec:loss-ablation} ablates the loss configuration choice.

\subsection{Data Strategies}
\label{sec:data}

Training on worldwide search logs introduces several biases that we address with targeted strategies:

\paragraph{Storefront Capping.}
Raw search logs are heavily skewed toward a small number of high-traffic storefronts. We cap the contribution of each storefront by downsampling any locale whose row count exceeds a percentile threshold across all storefronts, so that no single locale dominates the training distribution. The percentile threshold was selected by inspecting the frequency distribution across storefronts and identifying the point beyond which a small number of storefronts account for a disproportionate share of training data.

\paragraph{Entity-Type Capping.}
Songs dominate search engagement. We cap the contribution of each entity type by downsampling any type whose row count exceeds a percentile threshold, reducing the dominance of songs and ensuring that other entity types such as albums, artists, playlists, and radio stations retain adequate representation in training. This threshold was similarly chosen by inspecting the entity-type frequency distribution.

\paragraph{Exact-Match Downsampling.}
Queries that exactly match item metadata (e.g., searching ``Taylor Swift'' for the artist Taylor Swift) are downsampled to 20\% of their original frequency. Without this, the model collapses to lexical matching and fails to learn semantic generalization.

\paragraph{Synthetic Misspelling Augmentation.}
We augment 20\% of training queries with synthetic homophones (e.g., ``the beatles'' $\to$ ``the beetles'', ``payphone'' $\to$ ``payfone'') to improve robustness to the most common failure mode.

\subsection{Training Setup}

The model is trained on 8 NVIDIA A100 GPUs (40\,GB) with bf16 mixed-precision training. We use a batch size of 64 per device, a learning rate of $5 \times 10^{-5}$ with 1,000 warmup steps and a constant schedule thereafter, for one full epoch over the training data. The maximum context length is 512 tokens for documents and 32 tokens for queries. Music search queries are typically short (an artist name, song title, or a few descriptive words), so 32 tokens covers the vast majority of inputs; the 512-token document limit accommodates the structured metadata template along with lyric snippets. The production configuration emerged from approximately 100+ training iterations; Section~\ref{sec:ablations} presents ablation studies that justify the key design choices.

\section{Hybrid Retrieval}
\label{sec:hybrid}

\subsection{ANN Index}
\label{sec:ann}

Document embeddings are indexed using FAISS~\cite{johnson2019billion} with the composite configuration \texttt{OPQ128\_256,IVF2048\_HNSW32,PQ128} - optimized product quantization that learns a rotation of the 256-dimensional embedding space to align it with 128 sub-quantizers, an inverted file index with 2{,}048 coarse centroids whose traversal is accelerated by an HNSW graph (degree~32), and product quantization with 128 sub-quantizers for compressed storage.

Our catalog grows by hundreds of thousands of items per day, requiring frequent index refreshes to maintain retrieval freshness. We rebuild the ANN index every few hours using a batched pipeline. To control computational cost, we employ a change-detection mechanism that avoids recomputing embeddings for items whose underlying signals have not changed since the previous refresh. In practice, this significantly reduces embedding recomputation and enables low-latency index updates while maintaining high coverage of newly added or modified content.

\subsection{Score Blending}
\label{sec:blending}

A key challenge in hybrid retrieval is combining heterogeneous scoring signals that differ in both scale and distribution. In our system, downstream rankers are calibrated to a text-match score $s_{\text{text}}$ whose distribution is stable and tightly coupled to hand-tuned business logic. In contrast, cosine similarity scores $s_{\cos}$ from embedding retrieval are bounded in $[-1,1]$ and exhibit a markedly different distribution. Naively introducing $s_{\cos}$ as an additional feature would require retraining all downstream business logic and risk unintended shifts in ranking behavior.

To avoid this, we instead construct a single blended score:
\begin{equation}
    s_{\text{hybrid}} = \alpha \cdot s_{\text{text}} + \beta \cdot s_{\cos},
\end{equation}
where $\alpha$ and $\beta$ are scaling coefficients chosen to align the dynamic ranges of the two signals. In practice, $\alpha$ is fixed to $1.0$ and $\beta$ is tuned on a held-out validation set using the procedure described below, which provides a score-level calibration target.

\subsubsection{Training Signal via Spell-Correction Pairs}
\label{sec:training-signal}

Selecting $\beta$ purely by NDCG@K is under-specified: the objective is relatively flat near its optimum, so a wide range of $\beta$ values yield nearly identical ranking quality while placing substantially different weights on the embedding signal. To obtain a sharper, score-level target, we exploit \emph{spell-correction pairs} $(q, q^{*})$ mined from query logs, where $q$ contains a typo or morphological error and $q^{*}$ is its corrected form. For a relevant document $d$, $s_{\text{text}}(q, d)$ is depressed by lexical noise, whereas $s_{\text{text}}(q^{*}, d)$ reflects the target text-match score the system should ideally have produced on $q$. Since $s_{\cos}$ is largely invariant to such surface noise, $\beta$ should be just large enough to recover, in expectation, the clean-query score:
\begin{equation}
    s_{\text{hybrid}}(q, d) \;\approx\; s_{\text{text}}(q^{*}, d).
\end{equation}
We fit $\beta$ on a held-out set $\mathcal{H}$ via a one-dimensional search:
\begin{equation}
    \beta^{*} = \arg\min_{\beta}
    \sum_{(q, q^{*}) \in \mathcal{H}}
    \sum_{d \in \mathcal{D}_{q,q^{*}}}
    \ell\!\left(s_{\text{hybrid}}(q, d;\, \beta),\;
                s_{\text{text}}(q^{*}, d)\right),
\end{equation}
where $\mathcal{D}_{q,q^{*}}$ is a shared candidate set and $\ell$ is the Huber loss, chosen for robustness to imperfect correction pairs. Fitting $\beta$ at the score level also reduces the distributional distortion the quantile map $Q(\cdot)$, introduced next, must subsequently correct.

\subsubsection{Quantile Distribution Matching}
\label{sec:quantile-matching}

Even after linear blending, $s_{\text{hybrid}}$ does not follow the original text-match score distribution that downstream rankers implicitly assume. To address this, we apply \emph{quantile distribution matching} to map $s_{\text{hybrid}}$ back to the empirical distribution of $s_{\text{text}}$. Let $\{s_{p_i}\}$ and $\{f_{p_i}\}$ denote the percentile breakpoints of the hybrid and original text-match score distributions, respectively. We define a piecewise linear mapping:
\begin{equation}
    Q(s) = f_{p_i} + \frac{s - s_{p_i}}{s_{p_{i+1}} - s_{p_i}} \cdot (f_{p_{i+1}} - f_{p_i}), \quad s_{p_i} \leq s \leq s_{p_{i+1}}.
\end{equation}
This is a quantile-to-quantile mapping of the empirical CDF of $s_{\text{hybrid}}$ onto that of $s_{\text{text}}$: it preserves the relative ordering induced by $s_{\text{hybrid}}$ while aligning all percentile breakpoints between the two distributions.

\section{Ablation Studies}
\label{sec:ablations}

The ELISE encoder presented in Section~\ref{sec:model} emerged from over 100 training iterations spanning data composition, training objectives, backbone selection, loss functions, and architectural variants. This section distills the decisive findings that shaped the final design, including negative results that provide practical guidance for practitioners building similar systems.

\subsection{Data Quality Over Quantity}
\label{sec:data-ablation}

Training data for the QI objective was drawn from query logs, where a query led to a user engagement (play, add-to-library, or other conversion action). In addition to query logs, we explored augmenting with five supplementary sources: \emph{LLM-generated pairs} (query--item pairs produced via a doc2query approach~\cite{nogueira2019doc2query}, where an LLM generated plausible search queries from high-quality web pages describing catalog entities), \emph{web co-occurrence} (titles of web articles mentioning catalog entities, used as proxy queries), \emph{failed-query--item pairs} recovered from successful query rewrites, \emph{public playlists} paired with their constituent tracks, and \emph{editorial playlists} paired with their constituent tracks. Each source was intended to increase training coverage, particularly for tail queries underrepresented in engagement logs.

Contrary to expectation, removing \emph{any} supplementary source improved performance. Table~\ref{tab:data-ablation} reports a controlled small-scale ablation (${\sim}4\times$ fewer training steps than the full pipeline, otherwise identical model architecture and evaluation) isolating the effect of each source.

\begin{table}[t]
\caption{Data source ablation. Each row removes one supplementary source from the full training mix. \emph{Higher is better.}}
\label{tab:data-ablation}
\small
\begin{tabular}{lcc}
\toprule
\textbf{Training Data} & \textbf{Hit@10} & \textbf{$\Delta$} \\
\midrule
All sources                                  & 0.36 & --- \\
\; w/o LLM-generated pairs                   & 0.41 & $+$0.05 \\
\; w/o web co-occurrence                     & 0.38 & $+$0.02 \\
\; w/o failed-query--item pairs              & 0.37 & $+$0.01 \\
\; w/o public playlists                      & 0.38 & $+$0.02 \\
\; w/o editorial playlists                   & 0.37 & $+$0.01 \\
\bottomrule
\end{tabular}
\end{table}

Every supplementary source acts as a net distractor: removing any one of them improves Hit@10. Stripping away \emph{all} supplementary sources at once and training on raw query logs alone yields a $+0.06$ gain over the full mix, confirming that engagement-based query logs carry the dominant supervisory signal. The production model therefore uses exclusively raw query-log engagement pairs. We caution that these deltas should be read directionally: precise magnitudes depend on the mixing ratio between sources, which we did not exhaustively sweep.

Each supplementary source perturbs the query-to-engagement distribution that the QI objective targets, along one or both of two axes. \emph{Query distribution}: \emph{LLM-generated pairs}, \emph{web co-occurrence}, and the two \emph{playlist} sources supply proxy queries that are longer, grammatically complete, and skewed toward exploratory intents (e.g., ``upbeat workout music for morning runs''), whereas real queries are short, navigational, and noisy (e.g., ``bib dillan'', ``knockin heaven door''). \emph{Label fidelity}: the playlist sources pair a playlist title with its member tracks based on curator grouping rather than on user search-and-engagement behavior, and \emph{failed-query--item pairs} rely on rewrite-session proxies for items the original query did not engage---both yield weaker supervision than raw engagement. Query logs alone match both axes simultaneously, which is why they outperform any mix that includes the supplementary sources.

This contrasts with prior work showing LLM-generated training data can be effective for web retrieval~\cite{bonifacio2022inpars,dai2023promptagator}; the gap is plausibly larger in music search, where query patterns are highly domain-specific.

\subsection{Auxiliary Objectives}
\label{sec:aux-obj-ablation}

The primary QI objective trains the model to match queries to catalog items. We additionally explored two auxiliary objectives: Query--Query (QQ) and Item--Item (II), each designed to inject complementary signal into the shared embedding space.

The QQ objective aligns misspelled or reformulated queries with their corrected counterparts, using within-session query pairs where a user reformulated a failed query into a successful one. Prior work has shown that dual encoders are particularly vulnerable to misspelled queries~\cite{sidiropoulos2022misspellings}, and pre-training approaches have been proposed to improve robustness~\cite{zhuang2023typos}. Our QQ objective addresses this at fine-tuning time by explicitly training on misspelling-correction pairs. Table~\ref{tab:objective-ablation} shows the impact on Hit@10.

\begin{table}[t]
\caption{Training objective ablation.}
\label{tab:objective-ablation}
\small
\begin{tabular}{lcc}
\toprule
\textbf{Objectives} & \textbf{Hit@10} & \textbf{$\Delta$} \\
\midrule
QI only              & 0.54 & --- \\
QI + QQ              & 0.56 & +0.02 \\
QI + QQ + II         & 0.56 & +0.02 \\
\bottomrule
\end{tabular}
\end{table}

Adding QQ improved Hit@10 by +0.02, directly addressing misspelling and query reformulation failures. Adding II (item--item similarity, sourced from co-listens and playlist co-occurrence) did not further improve Hit@10. We hypothesize that the item--item signal is already implicitly captured by the QI objective: items that frequently co-occur in playlists or listening sessions tend to be engaged by the same queries, so the QI objective already pulls them into nearby regions of the embedding space.

The production model uses only QI + QQ. This finding suggests that auxiliary objectives are most valuable when they target a \emph{specific, measured failure mode}---in this case, QQ targets the misspelling and query reformulation failures that dominate non-converted sessions. The II objective adds complexity without addressing a concrete gap.

\subsection{Backbone Selection}
\label{sec:backbone-ablation}

The choice of pre-trained backbone proved to be the single highest-leverage decision. We evaluated two backbone families: E5-\linebreak multilingual-base and GTE-multilingual-base~\cite{li2023towards}. Table~\ref{tab:backbone} compares the backbones both before and after domain fine-tuning.

\begin{table}[t]
\caption{Backbone comparison. Base = off-the-shelf, no fine-tuning. Fine-tuned = best model from each backbone family after full training pipeline.}
\label{tab:backbone}
\small
\begin{tabular}{llcc}
\toprule
\textbf{Backbone} & \textbf{Config} & \textbf{Hit@10} & \textbf{$\Delta$} \\
\midrule
GTE-multilingual-base  & Base (256-d) & 0.39 & --- \\
Qwen3-Embedding-0.6B   & Base (256-d)         & 0.39 & --- \\
\midrule
E5-multilingual-base   & Fine-tuned (best)     & 0.56 & --- \\
GTE-multilingual-base  & Fine-tuned            & 0.61 & +0.05 \\
\bottomrule
\end{tabular}
\end{table}

Switching from E5-multilingual-base to GTE-multilingual-base yielded a +0.05 improvement in Hit@10---larger than any single change to training data, loss function, or objectives---owing to its more recent pre-training data pipeline covering 75+ languages, providing a stronger initialization for our retrieval task. Qwen3-Embedding-0.6B matched GTE's zero-shot performance (0.39) but was not pursued due to its larger parameter count.

\subsection{Loss Configuration}
\label{sec:loss-ablation}

The two-stage training schedule (Section~\ref{sec:curriculum}) combines a specific loss function with an implicit negative selection strategy---InfoNCE assigns gradient to every in-batch negative (softmax-weighted), while pairwise hinge with hardest-in-batch selection backprops through only the single hardest negative. We evaluate three loss configurations in a controlled setting where all other variables (data, backbone, training duration) are held constant. Table~\ref{tab:loss-ablation} reports results.

\begin{table}[t]
\caption{Loss configuration ablation. All models use GTE-multilingual-base with identical data and training setup; only the loss schedule differs.}
\label{tab:loss-ablation}
\small
\begin{tabular}{lc}
\toprule
\textbf{Loss Configuration} & \textbf{Hit@10} \\
\midrule
Two-stage hinge: all-neg hinge $\to$ hardest-neg hinge & 0.64 \\
Single-stage InfoNCE & 0.64 \\
Two-stage: InfoNCE $\to$ hardest-neg hinge (production) & \textbf{0.66} \\
\bottomrule
\end{tabular}
\end{table}

Neither InfoNCE alone nor a hinge-family two-stage schedule achieves the performance of the full configuration. InfoNCE alone provides gradient from many negatives but lacks the focused refinement of hardest-negative selection; the hinge-family curriculum already transitions from broad to focused negatives, but pairwise hinge produces zero gradient for negatives that satisfy the margin, limiting its ability to establish global embedding structure in early training. The InfoNCE-to-hinge configuration combines the strengths of both: InfoNCE's softmax ensures gradient flows from all negatives (including easy ones that anchor the embedding space), while the subsequent hardest-negative hinge stage sharpens boundaries for the most confusable pairs.

\subsection{Negative Results}
\label{sec:negative-results}

Table~\ref{tab:negative} summarizes approaches that degraded performance or provided insufficient improvement to justify their complexity.

\begin{table}[t]
\caption{Negative results. $\Delta$ is relative to the corresponding baseline model. All approaches were abandoned.}
\label{tab:negative}
\small
\begin{tabular}{p{2.6cm}cp{3.0cm}}
\toprule
\textbf{Approach} & \textbf{$\Delta$} & \textbf{Explanation} \\
\midrule
Audio fusion (soft attention over acoustic features) & $-$0.03 & Training queries are predominantly navigational; acoustic signal rarely helps disambiguate the intended item \\
\addlinespace
Layer freezing (6 / 9 layers) & $-$0.01 / $-$0.02 & Music-domain adaptation requires full-depth fine-tuning; degradation proportional to frozen layers \\
\addlinespace
Special tokens for template fields & $-$0.03 & Randomly initialized embeddings create optimization mismatch with pre-trained weights \\
\addlinespace
Hard example mining (2nd-phase fine-tuning) & $+$0.01 & Marginal gain did not justify pipeline complexity \\
\addlinespace
Selecting QI pairs by conversion share & $-$0.02 & Popularity bias: conversion share correlates with item popularity, skewing training toward head items \\
\addlinespace
Aggregated engagement data (replacing raw query logs) & $-$0.02 & Per-query normalization-and-expansion flattens the natural frequency distribution the QI objective relies on \\
\bottomrule
\end{tabular}
\end{table}

Two findings merit additional discussion. The \textbf{audio fusion} experiment added a late-fusion attention block on the item side that combined text representations with pre-extracted acoustic features (songs only; other entity types fell back to text). It regressed by $-0.03$ against the text-only control. We attribute this to the composition of our training data: the majority of search queries are navigational (titles, albums, artists) rather than content-oriented (genre, mood, audio descriptions), so acoustic signal rarely helps disambiguate the intended item. Cross-modal retrieval systems such as CLAP~\cite{elizalde2023clap} and MuLan~\cite{huang2022mulan} target content-based audio queries via paired audio-text pre-training, which was beyond the scope of our work.

The \textbf{hard example mining} pipeline identified query--item pairs that the current model scored poorly on, filtered them for relevance using an LLM, and interleaved the survivors with standard training data. The best result marginally exceeded the base model (+0.01), which did not justify the added pipeline complexity.

\section{Experiments}
\label{sec:experiments}

\subsection{Offline Evaluation}
\label{sec:offline}

\paragraph{Hit@10 Evaluation.}
We evaluate ELISE against the non-finetuned GTE-multilingual-base baseline on nine held-out sets, summarized in Table~\ref{tab:hit10}. Since exhaustive relevance annotations are infeasible for open-ended music queries---many catalog items may satisfy a single query---we adopt Hit@10: whether the known ground-truth item appears in the top 10 retrieved results. Hit@10 also served as our development criterion for tuning the training mix and hyperparameters. Seven sets (Similar Artists through Natural-Language Navigational) use LLM-generated queries with human-annotated ground truths; Misspellings uses search-log queries with human-annotated ground truths; and Edit-Distance Misspellings pairs catalog items with synthetic misspellings within a controlled edit distance.

\begin{table}[t]
\caption{Hit@10 across evaluation sets. Baseline is GTE-multilingual-base without fine-tuning. Relative improvement shown in parentheses.}
\label{tab:hit10}
\small
\begin{tabular}{lcc}
\toprule
\textbf{Evaluation Set} & \textbf{Baseline} & \textbf{ELISE} \\
\midrule
Similar Artists         & 0.02 & 0.25 \\
Activities              & 0.54 & 0.69\;(+28\%) \\
Decades                 & 0.49 & 0.71\;(+45\%) \\
Genre                   & 0.59 & 0.81\;(+37\%) \\
Mixed Category          & 0.43 & 0.65\;(+51\%) \\
Moods                   & 0.48 & 0.64\;(+33\%) \\
Natural-Language Navigational        & 0.30 & 0.85\;(+183\%) \\
Misspellings            & 0.42 & 0.79\;(+88\%) \\
Edit-Distance Missp.    & 0.26 & 0.53\;(+104\%) \\
\midrule
\textbf{Overall Average} & \textbf{0.39} & \textbf{0.66\;(+69\%)} \\
\bottomrule
\end{tabular}
\end{table}

ELISE achieves a 69\% overall improvement, with the largest gains on the hardest query types: navigational natural-language queries (+183\%), misspellings (+88\%), and similar-artist queries (0.02 to 0.25, a hard category where baseline performance is near zero).

\paragraph{Non-Converted Session Analysis.}
To understand how the model addresses specific failure modes, we evaluate on a small human-labeled dataset sampled from non-converted sessions in query logs, where each session was categorized by retrieval outcome (Table~\ref{tab:nonconv}).

\begin{table}[t]
\caption{Hit@10 on non-converted sessions by category.}
\label{tab:nonconv}
\small
\begin{tabular}{lcc}
\toprule
\textbf{Category} & \textbf{Baseline} & \textbf{ELISE} \\
\midrule
No results               & 0.29 & 0.71\;(+147\%) \\
Poor recall              & 0.08 & 0.61\;(+695\%) \\
Poor ranking             & 0.08 & 0.53\;(+538\%) \\
Good results (non-failure)   & 0.40 & 0.77\;(+94\%) \\
\midrule
\textbf{Overall} & \textbf{0.28} & \textbf{0.70\;(+150\%)} \\
\bottomrule
\end{tabular}
\end{table}

\paragraph{Production Quality Evaluation.}
To measure end-to-end search quality on the actual query distribution, we compare the hybrid retrieval system against the current production system using PL2B@10 (Position-weighted L2 norm, Boosted), a metric that captures both ranking quality and result availability in a single score:
\begin{equation}
\label{eq:pl2b}
\text{PL2B@}k = \sqrt{\frac{\sum_{i=1}^{k} \mathit{rel}_i^{2}\;\cdot\;\tfrac{1}{i}}{\sum_{i=1}^{k} \tfrac{1}{i}}},
\end{equation}
where $\mathit{rel}_i \in [0,1000]$ is a relevance label assigned to the result at position~$i$ by a combination of human and LLM judges, and $k$ is the evaluation depth. The position weighting $\tfrac{1}{i}$ discounts lower-ranked positions, while the ``boosted'' variant assigns a fixed score to empty positions (when fewer than $k$ results are returned), set between the ``Unacceptable'' and ``Acceptable'' relevance labels: this penalizes few-result queries (versus ignoring those positions) while still preferring an empty position over an irrelevant one, discouraging systems from padding with bad results.

We evaluate on a sample of queries from production search logs. On this set, the hybrid retrieval system achieves a +0.5\% relative improvement in average PL2B@10 and +0.6\% in query-frequency-weighted average PL2B@10 over the production baseline. These positive lifts confirm no regression in offline ranking quality; the larger tail-query gains appear in the online evaluation below.

\subsection{Online A/B Test}
\label{sec:ab}

We conducted a worldwide A/B test on the search results page across all storefronts over an 18-day period. The experiment was powered using a two-sided $t$-test with a minimum detectable effect (MDE) of 0.25\% relative change in conversion rate. While semantic retrieval improves recall, it can introduce noise into the results. We formulated our experiment to compare two strategies for controlling this noise, each receiving 10\% of traffic alongside a 10\% control:
\begin{itemize}
    \item \textbf{T1}: Semantic results included only if their text match score exceeds a threshold.
    \item \textbf{T2}: Semantic retrieval with a cap on the number of semantically retrieved results.
\end{itemize}
T2 emerged as the clear winner across all metrics and was selected as the launch candidate.

\paragraph{Key Metrics.}
Table~\ref{tab:ab-topline} summarizes the topline results for both treatments versus control.

\begin{table}[t]
\caption{A/B test topline metrics (T1 and T2 vs.\ Control). Confidence intervals and $p$-values are reported at significance threshold $\alpha = 0.04$. $^{*}$\,Statistically significant.}
\label{tab:ab-topline}
\small
\begin{tabular}{llccc}
\toprule
\textbf{Metric} & \textbf{Arm} & \textbf{Rel.\ $\Delta$} & \textbf{Rel.\ CI} & \textbf{$p$-value} \\
\midrule
\multirow{2}{*}{Conversion Rate} & T1 & $+0.81\%$ & --- & --- \\
                                  & T2 & $+2.28\%^{*}$ & $[+2.19,\;+2.37]\%$ & $<10^{-4}$ \\
\midrule
\multirow{2}{*}{No-Result Rate}  & T1 & $-6.98\%$ & --- & --- \\
                                  & T2 & $-86.0\%^{*}$ & $[-86.1,\;-85.8]\%$ & $<10^{-4}$ \\
\bottomrule
\end{tabular}
\end{table}

T2 outperforms T1 by a wide margin on both CR and no-result rate. T1's text-match threshold requirement filters out semantically relevant results that have no lexical overlap with the query---precisely the cases where semantic retrieval adds the most value, such as misspellings and natural-language queries. T2's count-based cap retains these results while still controlling noise, yielding a 2.8$\times$ larger CR lift and a 12$\times$ larger reduction in no-result rate.

\paragraph{Storefront Breakdown.}
All storefronts show positive CR lift, ranging from +0.5\% to +6.7\%, with no regressions observed in any market. The strongest gains appear in storefronts where cross-lingual and transliteration challenges are most acute, while CJK (Chinese, Japanese, Korean) storefronts show the smallest improvements, consistent with the limitations discussed in Section~\ref{sec:discussion}.

\paragraph{Query Frequency Analysis.}
Table~\ref{tab:ab-freq} segments the CR lift by query frequency. As expected, the largest improvement is in tail queries (+7.93\%), which are the primary target of semantic retrieval. Mid-frequency queries also benefit (+0.89\%), while head queries remain stable (+0.14\%), confirming that semantic retrieval does not degrade well-served popular queries.

\begin{table}[t]
\caption{CR lift by query frequency segment. Segment-level significance was not independently tested; the aggregate effect is significant (Table~\ref{tab:ab-topline}).}
\label{tab:ab-freq}
\small
\begin{tabular}{lrrc}
\toprule
\textbf{Segment} & \textbf{Unique Queries} & \textbf{Sessions} & \textbf{CR Lift (Rel)} \\
\midrule
Head & 0.6\%   & 39.8\% & +0.14\% \\
Mid  & 16.7\%  & 29.9\% & +0.89\% \\
Tail & 82.8\%  & 30.2\% & +7.93\% \\
\bottomrule
\end{tabular}
\end{table}

\paragraph{Item Discovery.}
Semantic search broadens the items surfaced for diverse queries. The average number of distinct queries per item increases by 4.7\% in the treatment, with the lift particularly pronounced for items that already attract a wide variety of queries, indicating that semantic retrieval enables items to be discovered through queries they would not have matched lexically.

\paragraph{Latency.}
End-to-end, the semantic retrieval path adds fewer than 55\,ms of latency at p95, well within the latency budget for this project.

\section{Discussion}
\label{sec:discussion}

We examine where our system succeeds, where it struggles, what our development process revealed, and the limitations of our findings.

\paragraph{Qualitative Analysis.}
The system addresses all four failure modes. The misspelled query ``gnarly by cat's eye'' now surfaces the correct artist; the Persian cross-lingual query ``{\arabicfont\rtl{مارتیک بهار}}'' retrieves ``Bahar'' by Martik via the aligned multilingual embedding space; for unavailable content, the system surfaces related items (e.g., other songs by the same artist) rather than empty results; and functional queries like ``party music from the 90s'' retrieve editorially curated playlists such as ``'90s Hits Essentials''.

\paragraph{CJK Storefronts.}
Gains are weakest in CJK storefronts: Japan (+1.4\%), China (+0.6\%), South Korea (+0.5\%). Logographic scripts, implicit word boundaries, and distinct transliteration conventions all pose challenges; improving CJK performance via specialized embedding models~\cite{xiao2023cpack} is a priority for future work.

\paragraph{Lessons from Model Development.}
Across 100+ training iterations, a consistent theme emerged: simpler approaches outperformed complex ones. Filtering to a single clean data source (query logs) beat augmenting with five additional sources; two training objectives (QI + QQ) matched the performance of three; and switching the pre-trained initialization delivered larger gains than any training-recipe change we evaluated. Consistent with recent findings in domain-specific retrieval~\cite{xiao2023cpack}, we hypothesize that prioritizing data quality and backbone selection before sophisticated training techniques is a productive ordering for similar systems---though validating this across domains and the broader landscape of multilingual encoders is left to future work.

\paragraph{Bias Mitigation and Limitations.}
Training on engagement logs introduces two principal biases: geographic skew (a few high-traffic storefronts dominate) and entity-type skew (songs dominate engagements). We mitigate both through percentile-based capping (Section~\ref{sec:data}). Empirically, the A/B test shows positive CR lift in every storefront, with the strongest gains in markets where cross-lingual and transliteration challenges are most acute. However, storefront-level capping is a coarse proxy: it does not address within-storefront variation (e.g., minority-language speakers in multilingual countries) or other fairness dimensions such as genre or artist popularity. Finer-grained debiasing and per-language evaluation are directions for future work.

\section{Conclusion}
\label{sec:conclusion}

We presented a curriculum-trained multilingual bi-encoder integrated into a
mature music search stack via quantile-calibrated hybrid retrieval, delivering
a tail-concentrated lift with no per-storefront regressions and an 86\%
reduction in no-result sessions. The deployed system is a starting point
rather than an endpoint: meaningful headroom remains in CJK storefronts, and
finer-grained debiasing and per-language evaluation are natural next steps.
Beyond search, the same embedding space is a candidate substrate for
voice-based query understanding, offline spell correction, and downstream
ranking signals. More broadly, we view the practical
recipe---a curriculum-trained multilingual bi-encoder, integrated via score
calibration rather than ranker rewrite---as a transferable pattern for
retrieval systems that need to improve recall in a mature stack without
disturbing it.

\begin{acks}
The authors thank Sean Suchter, Lauren Hauser, Sam Harami, Roberto Konow, Rohan Hajela, and Kartik Goyal for their contributions. Generative AI tools were used only for limited language polishing; all technical content, results, and final manuscript decisions were produced and verified by the authors.
\end{acks}

\bibliographystyle{ACM-Reference-Format}
\bibliography{references}

\end{document}